\documentclass[11pt]{article}

\usepackage{amsmath}
\usepackage{graphicx}
\usepackage{bm}
\usepackage{natbib}
\usepackage{algorithm}
\usepackage{algpseudocode}
\usepackage[parfill]{parskip}
\usepackage{xcolor}

\usepackage[margin=0.5in]{geometry}
\usepackage{hyperref} 

\title{Bayesian design and analysis of two-arm cluster randomised trials using assurance}
\author{Kevin J. Wilson \\
School of Mathematics, Statistics \& Physics, Newcastle University, UK}
\date{}

\begin{document}

\maketitle

\begin{abstract}
	
We consider the design of a two-arm superiority cluster randomised controlled trial (RCT) with a continuous outcome. We detail Bayesian inference for the analysis of the trial using a linear mixed-effects model. The treatment is compared to control using the posterior distribution for the treatment effect. We develop the form of the assurance to choose the sample size based on this analysis, and its evaluation using a two loop Monte Carlo sampling scheme. We assess the proposed approach, considering the effect of different forms of prior distribution, and the number of Monte Carlo samples needed in both loops for accurate determination of the assurance and sample size. Based on this assessment, we provide general advice on each of these choices. We apply the approach to the choice of sample size for a cluster RCT into post-stroke incontinence, and compare the resulting sample size to those from a power calculation and assurance based on a Wald test for the treatment effect. The Bayesian approach to design and analysis developed in this paper can offer advantages in terms of an increase in the robustness of the chosen sample size to parameter mis-specification and reduced sample sizes if prior information indicates the treatment effect is likely to be larger than the minimal clinically important difference.


\end{abstract}

\section{Introduction}

In a two-arm cluster randomised controlled trial (RCT) the randomisation to the treatment and control arms is performed at the cluster, rather than the individual, level. This might be logistically less challenging, for example as a result of the need to provide extensive additional training to staff in the treatment arm, or necessary, for example if the treatment requires specialist equipment which is restricted in number by cost or other factors. These advantages must be balanced against the need for a larger sample size to achieve the same statistical power, as a result of the dependence between trial outcomes for individuals in the same cluster.

The standard analysis of a cluster RCT is via a linear mixed-effects model incorporating a fixed treatment effect and random cluster effects. Inference is achieved using (restricted) maximum likelihood estimation. If we assume a Wald test for the treatment effect, then we can write down an approximation to the power function in closed form \citep{Eld06}. To use the power to choose a sample size requires estimates for the correlation between individuals within a cluster, the overall variability in the outcome variable and the variability in sample sizes between different clusters, as well as either the minimal clinically important difference (MCID) or an estimate for the treatment effect. If any (or all) of these parameters is mis-specified, then this will have an effect on the power function, resulting in an under- or over-powered trial. 

The power of the trial therefore is a conditional probability that the null hypothesis for the treatment effect is rejected, given the assumed treatment effect and the values chosen for the nuisance parameters. We will in general have uncertainty about the nuisance parameters. We can take this uncertainty into account in the sample size calculation by defining a prior distribution on the nuisance parameters and then integrating the power function over this uncertainty. The result is the probability that the treatment effect is significant conditional only on the MCID. If an estimate of the treatment effect is used instead of the MCID, then the prior distribution will also include our uncertainty on the treatment effect, and we obtain the uncontitional probability that the treatment effect will be significant following the trial. This latter quantity is known as the assurance in the literature \citep{Oha05,Kun21}, although we expand the definition to include the case where we still condition on the MCID for the treatment effect. Assurance has been proposed for cluster RCTs based on analysis via a Wald test and using an MCID on the treatment effect \citep{Wil22}.


In this paper, we propose and develop a Bayesian approach to choose a sample size for a two-arm cluster RCT with a continuous outcome, assumed to be normally distributed, when the final analysis of the trial is to be Bayesian. A systematic review on Bayesian methods in cluster RCTs \citep{Jon21} found thirteen papers which developed Bayesian methods. Five of these papers considered the design or analysis of cluster RCTs with binary or count outcomes (e.g., \cite{Cla10}). Another three were primarily concerned with estimation of the intra-cluster correlation coefficient (ICC, e.g., \cite{Tur06}). Of the remaining five, one developed a method for the analysis of rate ratios for repeated events, one was concerned with imputating a true endpoint from a surrogate and one modelled multivariate outcomes in hierarchical data. The remaining two papers, \cite{Kik10} and \cite{Spi01}, considered cluster RCTs with continuous responses.

Both \cite{Kik10} and \cite{Spi01} considered the analysis of a cluster RCT using the linear mixed-effects model which we will outline in Section \ref{sec:inf}. \cite{Kik10} then used an approach based on the expected net benefit to choose the sample size. This has computational advantages over the assurance, although would represent a comprehensive culture change in the way sample sizes are chosen in RCTs. \cite{Spi01} focused primarily on the inference problem, and considered suitable prior distributions for the parameters in the linear mixed-effects model and their impact on the posterior distribution. We investigate the effect of the choice of prior distribution in light of this in Section \ref{sec:comp}. \cite{Spi01} then detailed how to incorporate parameter uncertainty in both the ICC and the overall standard deviation into a power calculation for the required sample size. Thus, the approach taken to sample size choice was via assurance, as in \cite{Wil22}, but without incorporating uncertainty on the variability between the cluster sample sizes.

This paper extends the idea of assurance for cluster RCTs to the case where the final analysis is to be Bayesian. In contrast to the case where the analysis is to be via a hypothesis test \citep{Spi01,Wil22}, this results in a two loop Monte Carlo simulation scheme to evaluate the assurance, based on posterior inference via Markov chain Monte Carlo (MCMC) methods. We outline the linear mixed-effects model to analysis a cluster RCT, and posterior inference via MCMC, in Section \ref{sec:inf}. We then describe our general approach to choose the sample size using assurance, incorporating the MCMC posterior samples, in Section \ref{sec:ss}. 
Thus we will see that using the assurance with a Bayesian analysis of the trial is a special case of Bayesian design of experiments \citep{Rya15}, where the utility is given by an indicator function describing whether the trial achieves a successful outcome. 

We investigate the effect of the choice of prior distribution for the analysis on the resulting inference and the number of Monte Carlo samples required, in both loops of the Monte Carlo scheme, for accurate assessment of the sample size and assurance, in Section \ref{sec:comp}, and provide general advice on suitable choices. We apply our Bayesian approach to the choice of sample size for cluster RCTs to an example trial, ICONS, in Section \ref{sec:app}, comparing the resulting sample sizes to those chosen using power and assurance based on a frequentist analysis of the trial. We see that, if we have a relatively high probability, based on our prior information, that the treatment effect will be larger than the MCID, then our proposed Bayesian approach has the potential to reduce the sample size required for the cluster RCT. More generally, the consideration of prior uncertainty on each of the parameters required for the sample size calculation results in a sample size which is more robust to mis-specification.

\section{Inference for a two-arm superiority cluster RCT}
\label{sec:inf}

Consider a two-arm superiortity cluster RCT comparing treatment with control. Observations of the primary outcome in the trial take the form $Y_{ij}$ for individual $i=1,\ldots,n_j$ in cluster $j=1,\ldots,J$. Note that there are typically different numbers of individuals in different clusters (i.e., $n_j\neq n_{j^{'}}$ in general). A standard analysis considers a linear mixed-effects model of the form $Y_{ij}\sim\textrm{N}(\mu_{ij},\sigma^2_w)$, where the linear predictor is given by
\begin{displaymath}
\mu_{ij} = \lambda + X_j\delta + c_j.
\end{displaymath}
In this model, $X_j=1$ if cluster $j$ is in the treatment arm and $0$ otherwise, $\lambda$ is an overall mean effect, $\delta$ is the treatment effect and $c_j\sim\textrm{N}(0,\sigma^2_b)$ is a random cluster effect, with $\sigma^2_b$ representing the between cluster variability. To perform Bayesian inference we give prior distributions to the parameters $\bm\theta = (\lambda,\delta,\sigma^2_b,\sigma^2_w)^{'}$. Suitable forms for the marginal prior distributions are
\begin{eqnarray*}
\lambda \sim \textrm{N}(m_\lambda,v_\lambda), &&
\delta \sim \textrm{N}(m_\delta,v_\delta), \\
\tau_b = \dfrac{1}{\sigma^2_b}  \sim  \textrm{Gamma}(r_b,s_b), &&
\tau_w = \dfrac{1}{\sigma^2_w}  \sim  \textrm{Gamma}(r_w,s_w).
\end{eqnarray*}
The prior distributions on $(\sigma^2_b,\sigma^2_w)$ represent cases (a) and (b) from \cite{Spi01}. Alternative forms are provided in cases (c)-(g) of \cite{Spi01} and \cite{Wil22}. We will investigate the effect of different prior distributions for $(\sigma^2_b,\sigma^2_w)$ on the resulting inference in Section \ref{sec:comp}.

For convenience in developing our methods later, we collect these marginal prior distributions together in the ``analysis prior'', whose probability density function (pdf) is denoted $\pi_A(\bm\theta)$. Given the relative complexity of the model, conjugate inference is not possible to obtain the posterior distribution. However, linear mixed-effects models are amenable to common Markov chain Monte Carlo (MCMC) approaches, and so we can easily obtain samples from the analysis posterior distribution of the parameters $\pi_A(\bm\theta\mid\bm y)$, where $\bm y = (y_{11},\ldots,y_{n_11},\ldots,y_{n_JJ})^{'}$. In this paper we will use the rjags package \citep{Plu22} for this task.

The decision on the superiorirty, or not, of the treatment to control, would then be based on the analysis posterior distribution for the treatment effect, $\pi_A(\delta\mid\bm y)$. In particular, we would declare the treatment to be superior if the probability that this effect is positive is large, given the data, i.e., 
$\Pr{}_A(\delta>0\mid\bm y)\geq \gamma$,
where standard values are $\gamma=0.95,0.99,0.999$. Again, this could be evaluated approximately based on MCMC samples from the posterior distribution for $\delta$. If a particular minimum effect size $\delta^{*}>0$ was of interest, then we could alternatively consider the posterior probability $\Pr{}_A(\delta>\delta^{*}\mid\bm y)\geq \gamma$ as the object of the inference.

\section{Sample size determination}
\label{sec:ss}

A critical task in the planning of any cluster RCT is the choice of a suitable sample size. This ensures that sufficient individuals are recruited to determine if the treatment is superior to control, with reasonable probability. Assurance is the Bayesian alternative to power for this task. 

In a cluster RCT the total sample size is given by $n_T = \sum_{j=1}^J n_j.$ The standard approach is to specify a total sample size, rather than the individual sample sizes for each cluster. Typically the number of clusters $J$ is either set in advance or a range of possible cluster sizes is considered. We can think of the number of individuals in each cluster, $\bm n = (n_1,\ldots,n_J)^{'}$, as coming from a multinomial distribution
$\bm n\sim\textrm{Multinomial}(n_T,\bm p),$ 
where $\bm p = (p_1,\ldots,p_J)^{'}$ and $p_j$ is the probability a randomly selected individual comes from cluster $j$. We could give a prior distribution to $\bm p$ which represents a prior belief that, while we expect variation between the cluster sizes, {\em a priori} we do not know which cluster sizes are likely to be larger or smaller. To do so we could choose a symmetrical Dirichlet prior
$\bm p \sim \textrm{Dirichlet}(\bm a),$
where $\bm a=(a_1,\ldots,a_J)^{'}$, and $a = a_1=a_2=\ldots=a_J$. Smaller values of $a$ will give more variation in cluster sizes and larger values will give less variation.

Define the event $S$ to be the trial resulting in a successful outcome, i.e., concluding that the treament is superior to control. The assurance in this case, for sample size $n_T$, is given by
\begin{eqnarray*}
A(n_T) & = & \int\Pr{}_A(S\mid\bm\psi,n_T)\pi_D(\bm\psi)d\bm\psi \\
& = & \int\int I_A[S\mid\bm z] f(\bm z\mid\bm\psi,n_T)\pi_D(\bm\psi)d\bm\psi d\bm z,
\end{eqnarray*}
where $\bm z=(\bm y, \bm n)^{'}$ is a vector containing all observations and $\bm\psi=(\bm\theta,\bm p)^{'}$ is a vector containing all parameters relevant to the design of the trial, $I_A[S\mid\bm z]$ is an indicator function taking the value 1 if the trial is a success and 0 if not, given observations $\bm z$ and analysis prior distribution $\pi_A(\bm\theta)$, $f(\bm z\mid\bm\psi,n_T)$ is the pdf of $\bm z$ and $\pi_D(\bm\psi)$ is the design prior distribution for $\bm\psi$. The design prior distribution is the prior used to calculate the sample size for the trial, as opposed to the analysis prior distribution, which is used for the analysis at the end of the trial. Typically we would wish to use the best information we have about the model parameters when choosing the sample size, whereas we may wish the inference following the trial to be dominated by the data, necessitating a much less informative prior distribution at that stage. For a comprehensive discussion of design and analysis priors see Section 3 of \cite{Oha01}.

We can express the latter two terms as, respectively,
\begin{eqnarray*}
f(\bm z\mid\bm\psi,n_T) & = & f(\bm y\mid\bm\theta,\bm n)f(\bm n\mid\bm p, n_T) \\
\pi_D(\bm\psi) & = & \pi_D(\bm\theta)\pi_D(\bm p),
\end{eqnarray*}
where $f(\bm y\mid\bm \theta,\bm n)$ is the pdf of $\bm y$, $f(\bm n\mid\bm p, n_T)$ is the probability mass function of $\bm n$ and $\pi_D(\bm p)$ is the design prior for $\bm p$. This means that the integral over $\bm z$ above is actually an integral over $\bm y$ and a sum over $\bm n$, but we choose to leave it in its simpler form. In the case where we will declare the trial a success based on the probability, in the analysis posterior distribution, that the treatment effect is positive, the term $I_A[S\mid\bm z]$ will take the form
$I_A[S\mid\bm z] = I\left[\Pr{}_A(\delta>0\mid\bm y)>\gamma\right].$

A standard approach to evaluate $A(n_T)$ is via simulation. As in Bayesian design of experiments problems generally, the simulation would typically take a two loop structure; with an ``outer loop'' and an ``inner loop''. In the outer loop we would take samples $\bm\psi^{(\ell)}$ for $\ell=1,\ldots,L$ from their design prior distribution, and then sample $\bm z^{(\ell)}$ from their probability density/mass functions based on these values. This would allow us to obtain an approximation to the assurance, for sample size $n_T$, as
\begin{displaymath}
\tilde{A}(n_T) = \dfrac{1}{L}\sum_{\ell=1}^{L} I_A\left[S\mid \bm z^{(\ell)}\right]
\end{displaymath} 
For each outer loop sample, in the inner loop we would take an approximation to $I_A[S\mid \bm z^{(\ell)}]$ by sampling $\delta^{(k)}$ for $k=1,\ldots,K$, via MCMC, from the analysis posterior distribution for $\delta$ given observations $\bm z^{(\ell)}$. This gives the approximation
\begin{eqnarray*}
\tilde{I}_A[S\mid\bm z^{(\ell)}] & = & \tilde{\Pr}{}_A(\delta>0\mid\bm y^{(\ell)}) \\
& = & \dfrac{1}{K}\sum_{k=1}^KI\left[\delta^{(k)}>0\right].
\end{eqnarray*}
That is, we have $L$ samples in the outer loop and, for each, we require $K$ samples in the inner loop, making $L\times K$ samples in total, excluding burn-in, thinning etc. in the MCMC chains. This will provide the assurance for a chosen sample size. To find the minimum sample size which gives a particular level of assurance, say 80\%, this process can be repeated for increasing sample sizes until the required sample size is found. Further repetition of the entire process would allow us to find the required sample sizes for different numbers of clusters as required. 

Given that, for each step, $L$ sets of simulated values are taken and, for each, a full MCMC run is needed, then the process has the potential to become computationally intensive and, possibly, slow to evaluate. We will investigate how many Monte Carlo samples in the outer loop and MCMC samples in the inner loop are required to provide accurate sample sizes in Section \ref{sec:comp}. However, waiting for minutes, or even hours, to find the required sample size for a clinical trial which will last for several years is not in itself a problem.

\section{Investigating the analysis prior and Monte Carlo error}
\label{sec:comp}

\subsection{The effect of the analysis prior}

In this section we will evaluate the effect of different choices of analysis prior on the Bayesian inference for the treatment effect, $\delta$, the object of the inference for the sample size calculation. To do so, we will use 100,000 MCMC samples, following burn-in, to evaluate posterior distributions, so that the posterior distributions from the MCMC samples are almost exact. 

We consider inference based on a single set of observations sampled from the model. We set the within group variance to be half of the between group variance $\sigma^2_w=1, \sigma^2_b=2$, the overall mean $\lambda=10$, a positive treatment effect of $\delta=1$, and use equal probabilities of an individual belonging to each cluster. We set the average number of individuals per cluster to be 10 and will consider the effect of varying the number of clusters (and hence total sample size). We choose diffuse analysis priors for the parameters in each case, representing a lack of knowledge, or an unwillingness to impose beliefs into the prior. Using the forms of the prior distributions in Section \ref{sec:inf}, we take $m_\lambda=1, m_\delta=0, v_\lambda=v_\delta=1000$, and $r_b=r_w=s_b=s_w=0.001$. We observe that the priors are not centred on the true parameter values.

\cite{Spi01} found that the forms of the marginal analysis prior distributions for $(\sigma^2_b,\sigma^2_w)$ can have an effect on the resulting inference, even when the priors are chosen to be non-informative, as is the case above. To see how strong this effect is for our linear mixed-effects model and typical sample sizes, we investigate inference for the analysis priors described above and five further sets of analysis priors for $(\sigma^2_b,\sigma^2_w)$ based on those in \ref{Spi01}. The 6 sets of analysis prior distributions considered are given in Table \ref{tab:analysis}. We see that the priors 1. and 2. are of the same form, but with different values for the hyper-parameters. Each of the other priors gives $\sigma^2_w$ a log-uniform prior distribution. In priors 3. and 4. $\sigma^2_b$ is given a uniform prior distribution and a log-uniform prior respectively. Priors 5. and 6. are parameterised in terms of the ICC, $\rho=\sigma^2_b/(\sigma^2_b+\sigma^2_w)$, giving it a uniform and beta prior distribution respectively. This implies a prior distribution for $\sigma^2_b$.

\begin{table}[ht]
\centering
\begin{tabular}{|c|cc|}\hline
Prior number & Prior Distributions & Parameter values \\ \hline
1 & $1/\sigma^2_b  \sim  \textrm{Gamma}(r_b,s_b)$ & $r_b=s_b=0.001$ \\
& $1/\sigma^2_w  \sim  \textrm{Gamma}(r_w,s_w)$ & $r_w=s_w=0.001$ \\ \hline
2 & $1/\sigma^2_b  \sim  \textrm{Gamma}(r_b,s_b)$ & $r_b=s_b=0.1$ \\
& $1/\sigma^2_w  \sim  \textrm{Gamma}(r_w,s_w)$ & $r_w=s_w=0.1$ \\ \hline
3 & $\log(\sigma^2_b) \sim \textrm{Unif}(l_b,u_b)$ & $l_b=-10,u_b=10$ \\
& $\log(\sigma^2_w) \sim \textrm{Unif}(l_w,u_w)$ & $l_w=-10,u_w=10$ \\ \hline
4 & $\sigma^2_b \sim \textrm{Unif}(l_b,u_b)$ & $l_b=0,u_b=100$ \\
& $\log(\sigma^2_w) \sim \textrm{Unif}(l_w,u_w)$ & $l_w=-10,u_w=10$  \\ \hline
5 & $\rho \sim \textrm{Unif}(l_\rho,u_\rho)$ & $l_\rho=0,u_\rho=1$ \\
& $\log(\sigma^2_w) \sim \textrm{Unif}(l_w,u_w)$ & $l_w=-10,u_w=10$  \\ \hline
6 & $\rho \sim \textrm{Beta}(r_\rho,s_\rho)$ & $r_\rho=1,s_\rho=1$ \\
& $\log(\sigma^2_w) \sim \textrm{Unif}(l_w,u_w)$ & $l_w=-10,u_w=10$  \\ \hline
\end{tabular}
\caption{The 6 different sets of analysis prior distributions considered for $(\sigma^2_b,\sigma^2_w)$.}
\label{tab:analysis}
\end{table}

\begin{figure}[h!]
\centering
\includegraphics[width=0.24\linewidth]{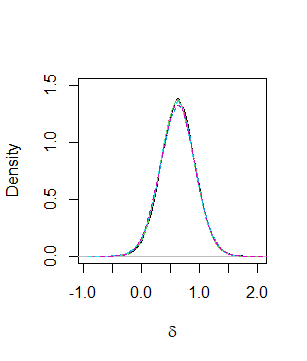}
\includegraphics[width=0.24\linewidth]{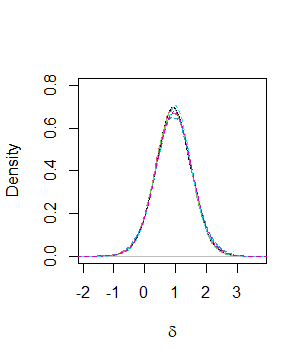}
\includegraphics[width=0.24\linewidth]{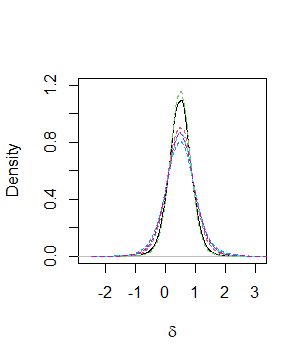}
\includegraphics[width=0.24\linewidth]{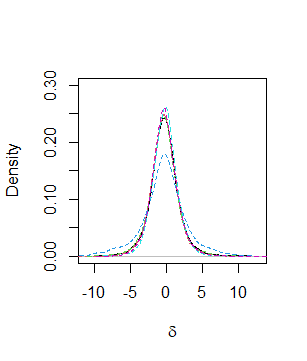}
\includegraphics[width=0.24\linewidth]{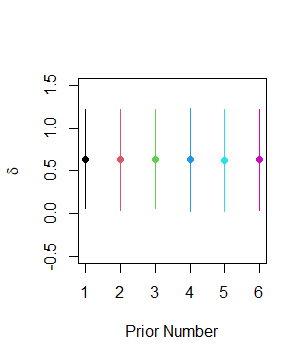}
\includegraphics[width=0.24\linewidth]{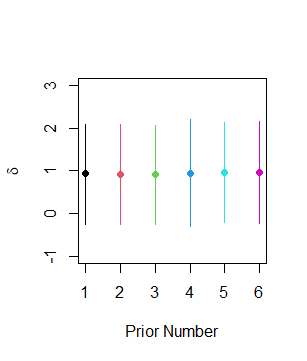}
\includegraphics[width=0.24\linewidth]{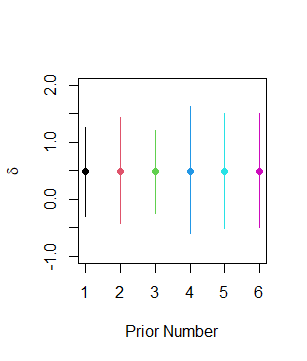}
\includegraphics[width=0.24\linewidth]{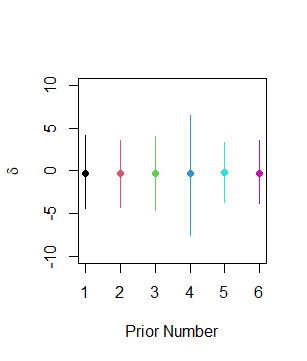}
\caption{The posterior densities (top) and posterior means and symmetric 95\% intervals (right) of $\delta$ for sample sizes of 500 (top left), 250 (top right), 120 (bottom left) and 60 (bottom right) based on priors 1-6, respectively in black, red, green, blue, cyan and magenta.}
\label{fig:post}
\end{figure}

The posterior distributions for the  treatment effect, based on 100,000 samples, are provided in the top panel of Figure \ref{fig:post}. We consider $J=(6,12,25,50)$ clusters giving total sample sizes of $n_T=(60,120,250,500)$. 

We see that for sample sizes of 250 and 500, all priors give essentially identical inference for $\delta$. For the smaller sample sizes of 120 and particularly 60, all of the priors give very similar posterior inference for the mean of $\delta$, although there are some differences in the variability. In particular, for a sample size of 60, analysis prior 4. shows substantially more posterior variability in $\delta$ than the other 5 prior distributions. This can be seen more clearly in the bottom panel of Figure \ref{fig:post}. We repeated this comparison, varying the values of $\delta,\lambda$ and $\sigma^2_b/\sigma^2_w$. The results are provided in Section A of the supplementary material. The messages are consistent with the case described above, with the majority of the analysis priors giving very similar inference, with the exception of prior 4.

The sample size, and the success or otherwise of the RCT, will be based on the posterior probability that $\delta>0$. We provide this for each set of analysis priors in Table \ref{tab:delta}, alongside the time taken to obtain the posterior samples of $\delta$, for a sample size of 500 on a laptop with an Intel(R) Core(TM) i5-8350U CPU @ 1.70GHz, 1896 Mhz, 4 Cores, 8 Logical Processors.

\begin{table}[ht]
	\centering
	\begin{tabular}{|c|cccc|c|}\hline
		Prior number & \multicolumn{4}{|c|}{Sample size} & Time (s) \\
	& 500 & 250 & 120 & 60 & \\ \hline
	1 & 0.98 & 0.94 & 0.90 & 0.43 & 7.64 \\
	2 & 0.98 & 0.94 & 0.87 & 0.43 & 7.09 \\
	3 & 0.98 & 0.94 & 0.91 & 0.41 & 50.76 \\
	4 & 0.98 & 0.94 & 0.83 & 0.45 & 46.64 \\
	5 & 0.98 & 0.95 & 0.85 & 0.45 & 54.00 \\
	6 & 0.98 & 0.94 & 0.85 & 0.42 & 53.28 \\ \hline
	\end{tabular}
	\caption{The posterior probability that $\delta>0$ for each of the different sample sizes considered, for analysis priors 1-6 for $(\sigma^2_b,\sigma^2_w)$. Also included is the time taken to obtain the posterior samples of $\delta$ when the sample size is 500.}
	\label{tab:delta}
\end{table}

We see that the posterior probabilities are very similar for all methods and for all sample sizes, with the exception of analysis prior 4., which provides lower posterior probabilities than the others, particularly for smaller sample sizes. We also see that analysis priors 1. and 2. offer substantially faster inference than priors 3-6, as a result of being the conjugate forms for $\sigma^2_b,\sigma^2_w$. Given the increase in speed compared to the other priors, and the equivalent posterior inference to the majority of the other priors, we choose to use analysis priors 2. for all future analyses in this paper. 

\subsection{The effect of sampling}

We now consider the assurance and sample sizes resulting from the posterior inference using analysis prior 2. In particular, we investigate the number of Monte Carlo samples, $L$, required in the outer loop and how many MCMC samples, $K$, are required in the inner loop, to get accurate values for the assurance and sample size. We choose Normal design priors centred on the truth, with variances of 1 for $\lambda$ and $\delta$, 0.01 for $\sigma^2_w$ and 0.04 for $\sigma^2_b$. We use a Dirichlet prior for $p$ with $a=100$. To allow the approach to run in reasonable time, we consider designs with $J=10$ clusters. 

\begin{figure}[ht]
\centering
\includegraphics[height=7.5cm]{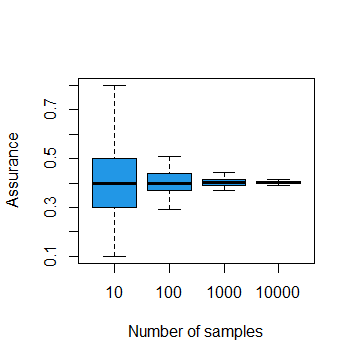}
\includegraphics[height=7.5cm]{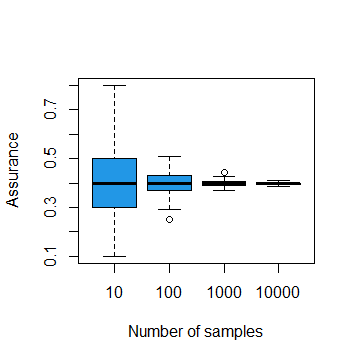}
\includegraphics[height=8cm]{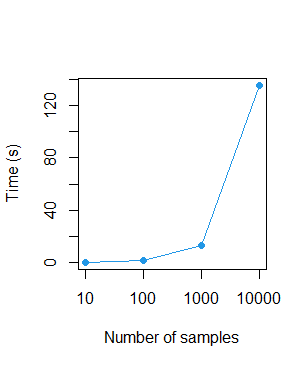}
\includegraphics[height=8cm]{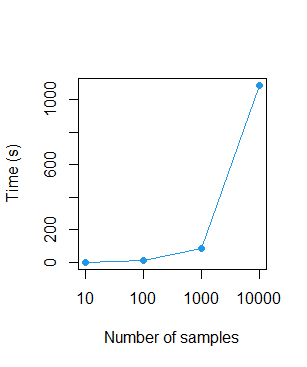}
\caption{Top: boxplots of the assurance from 100 repetitions, based on the number of MC samples in the outer loop being 10, 100, 1000 and 10,000 respectively for 1000 (left) and 10,000 (right) MCMC samples. Bottom: the time to compute the assurance for each combination.}	
\label{fig:assMC}
\end{figure}

We choose an average sample size per cluster of $\bar{n}=5$. Note that approximately 95\% of the design prior density for $\delta$ is above zero, and so we can achieve an assurance of up to 95\% in this case. We consider numbers of inner MCMC samples in the range $K\in[10,10000]$ and investigate the effect of varying the number of outer loop samples $L\in[10,10000]$. We repeat the assurance calculation 100 times for each combination of $L$ and $K$ and provide boxplots of the results in the top row of Figure \ref{fig:assMC}.
We see that, as expected, the variability in the estimate of the assurance drops with increasing number of MC samples. There appears to be a smaller effect of the number of inner loop samples, between 1000 and 10,000. It appears that 1000 or 10,000 samples in the outer loop alongside either 1000 or 10,000 samples in the inner loop may provide an accurate value of the assurance. In addition, we recorded the time to calculate the assurance in each case, which is given on the log-scale in the bottom row of Figure \ref{fig:assMC}. We re-ran this analysis, varying the values of $\delta,\lambda$ and $\sigma^2_b/\sigma^2_w$. A summary of the results is provided in Section B of the supplementary material. For all combinations of parameters considered, the behaviour of the assurance estimates with changing numbers of samples is consistent with that observed in Figure \ref{fig:assMC}.  

To assess whether the approximations to the assurance are sufficiently accurate, and to allow us to provide some advice on how many outer and inner loop samples may be sufficient, we also consider the sample size resulting from the assurance calculation. We set an assurance target of 0.8. For each combination of $L,K\in\{1000, 10000\}$ we find the sample size 100 times and report the modal sample size together with the proportion of occasions on which this sample size was found in Table \ref{tab:modss}. We also report the number of repetitions needed to ensure the correct sample size is chosen as the modal sample size with probability at least 90\% and the time taken to calculate the sample size a single time. 

\begin{table}[ht]
\centering
\begin{tabular}{|cc|cc|cc|} \hline
$K$ & $L$ & Mode & Proportion & Repetitions & Time (s) \\ \hline
1000 & 1000 & 80 & 0.55 & 35 & 62 \\
10,000 & 1000 & 80 & 0.60 & 16 & 558 \\
1000 & 10,000 &  80  & 0.80  & 5  & 705 \\
10,000 & 10,000 & 80  & 0.95  &  1  & 5156 \\
 \hline
\end{tabular}
\caption{The modal sample sizes and proportion of occasions on which they were recorded, based on 100 simulations, for number of outer loop samples of $1000$ and 10,000 and number of inner loop samples of 1000 and 10,000.}
\label{tab:modss}
\end{table}

We see that all combinations of $L,K$ obtain the correct sample size the majority of the time. To be confident in the modal sample size requires between 1 and 35 repetitions depending on the values of $L,K$. In this case, the fastest strategy would be to use $L=K=1000$ and repeat the sample size calculation 35 times, choosing the modal sample size. This would take just over half an hour. We will use these specifications when calculating sample sizes for the application in the next section.

\section{Application: the ICONS study}
\label{sec:app}

\subsection{The ICONS study}

In this section we will apply the Bayesian approach to calculate an appropriate sample size for the ICONS cluster RCT \citep{Tho15}.The trial considers the effectiveness of a systematic voiding programme for people admitted to NHS stroke units with urinary incontinence. The primary outcome is the incontinence symptom severity total score at 3 months post-randomisation. The voiding programme was compared to usual care. The orinigal sample size was chosen based on a power calculation assuming a two-sample t-test on the primary outcome. The power calculation uses a slightly different parameterisation, via the overall variance, $\sigma^2=\sigma^2_b + \sigma^2_w$, and the ICC, $\rho=\sigma^2_b/(\sigma^2_b+\sigma^2_w)$, in place of $(\sigma^2_b,\sigma^2_w)$. The variability between cluster sizes is represented by the coefficient of variation in cluster sizes, denoted $\nu$. In the ICONS power calculation, the MCID was assumed to be $\delta_M=2.52$, the overall standard deviation was estimated to be $\hat{\sigma}^2=8.32$, the ICC was estimated to be $\hat{\rho}=0.028$ and the number of clusters considered was 40-50 (so 20-25 in each arm). The estimates were based on a feasibility study \citep{Tho15}. We consider a one-tailed test for comparison to the Bayesian approach, although the original calculation was performed for a two-tailed test. We use a significance level of $\alpha=0.05$ and a desired power of $1-\beta=0.8$.

In this case, an approximation to the power function is given by \citep{Eld06,Wil22}
\begin{equation} \label{eq:power}
P(n_T\mid\delta_M,\hat{\sigma},\hat{\rho},\hat{\nu}) = \Phi\left(\sqrt{\dfrac{J\bar{n}}{4[1+\{(\hat{\nu}^2+1)\bar{n}-1\}\hat{\rho}]\hat{\sigma}^2}}\delta_M-z_{1-\alpha}\right),
\end{equation}
where $\bar{n}$ is the average sample size per cluster and $z_{1-\alpha}$ is the $100(1-\alpha)\%$ percentile of the standard normal distribution. Using this, we can calculate the required sample sizes for total numbers of clusters between 40 and 50. They are given in the top row of Table \ref{tab:ss}. We see that between 329 (47 clusters) and 378 (42 clusters) individuals would need to be recruited to the trial. Note that the sample sizes can increase locally for increasing numbers of clusters when the average number of individuals required per cluster does not drop with an increase in the number of clusters. 

\begin{table}[ht]
\centering
\begin{tabular}{|c|ccccccccccc|} \hline
 & \multicolumn{11}{c|}{Number of clusters} \\ 
 & 40 & 41 & 42 & 43 & 44 & 45 & 46 & 47 & 48 & 49 & 50 \\ \hline
 Power &  360 & 369 & \textcolor{red}{378} & 344 & 352 & 360 & 368 & \textcolor{blue}{329} & 336 & 343 & 350 \\
 Hybrid MCID & \textcolor{red}{480} & 451 & 462 & 430 & 440 & 450 & 414 & 423 & 432 & 441 & \textcolor{blue}{400} \\
 Full hybrid & 240 & 246 & \textcolor{red}{252} & \textcolor{blue}{215} & 220 & 225 & 230 & 235 & 240 & 245 & 200 \\
 Bayes & \textcolor{red}{240} & 205 & 210 & 215 & 220 & 225 & 230 & 235 & \textcolor{blue}{192} & 196 & 200 \\ \hline
\end{tabular}
\caption{Total sample sizes required using power, hybrid assurance and assurance using a Bayesian analysis, based on having between 40 and 50 clusters. The smallest sample size for each approach is given in blue and the largest sample size is given in red.}
\label{tab:ss}
\end{table}

\subsection{A hybrid approach}

We compare the power calculation to the hybrid approach from \cite{Wil22}. In the hybrid approach a design prior distribution is defined over the unknown parameters and the assurance is given by the power averaged over this prior distribution. In practice, we find the power using (\ref{eq:power}) for a set of samples from the design prior distribution, and then calculate the average power. This is a hybrid version of the assurance. We consider two versions of this hybrid assurance. The first considers a design prior distribution on $(\sigma,\rho,\nu)$ and the MCID $\delta_M$ and the second considers a design prior distribution over $(\delta,\sigma,\rho,\nu)$. We name these the ``hybrid MCID'' and ``full hybrid'' approaches respectively.

We use the design prior distributions given in \cite{Wil22}. That is, 
based on the data from the ICONS feasibility study, we give $\sigma$ and $\nu$ gamma marginal prior distributions, centred at $m_\sigma=8.32$ and $m_\nu=0.49$, respectively, and with variances of $v_\sigma=1^2$ and $v_\nu=0.066^2$. The prior distribution for $\rho$ is based on Bayesian hierarchical modelling combining 34 ICC estimates extracted from 16 previous similar trials. Full details are provided in \cite{Tis22}. We use the resulting MCMC samples from this model as samples from the design prior distribution for $\rho$. We define a joint prior distribution between $\rho$ and $\sigma$ via a Gaussian copula with correlation parameter $0.44$. For the full hybrid method we also need a design prior distribution for $\delta$. We will consider the effect of this prior in the next section, but will initially set this to be $\delta\sim\textrm{N}(3.5,0.9^2)$, which represents a belief that the treatment effect is very likely to be in the range $[1,6]$ and has a probability of 0.75 of being above the MCID of 2.52.

We can visualise the marginal design prior distributions of each of $(\delta,\sigma,\rho,\nu)$ in Figure \ref{fig:priors}. A plot of the joint prior distribution for $(\rho,\sigma)$ is provided in Section C of the supplementary material. We plot the empirical distributions of each based on 10,000 samples.
\begin{figure}[ht]
\centering
\includegraphics[height=3in]{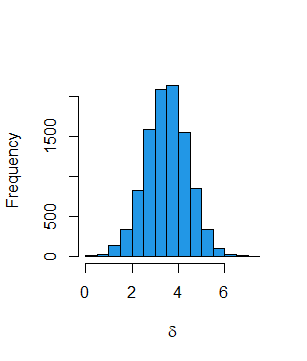}
\includegraphics[height=3in]{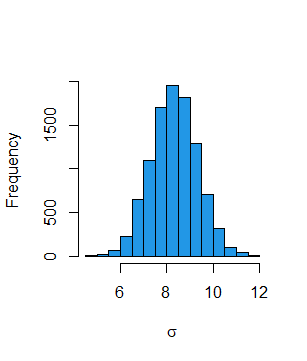}
\includegraphics[height=3in]{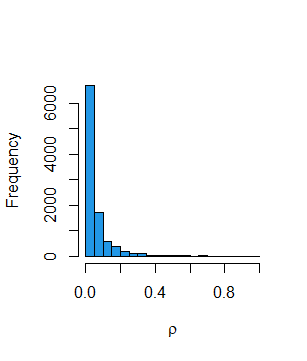}
\includegraphics[height=3in]{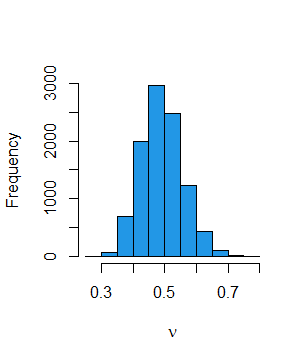}
\caption{10,000 samples from the marginal design prior distributions of $\delta$ (top left), $\sigma$ (top right), $\rho$ (bottom left) and $\nu$ (bottom right).}
\label{fig:priors}
\end{figure}

Based on these priors, we calculate the sample sizes required for the hybrid MCID and full hybrid methods, and report them in the second and third rows of Table \ref{tab:ss}. We see that the hybrid MCID approach requires larger sample sizes than the power calculation, as a result of the uncertainty in the estimates of the model parameters induced by the prior distribution. However, the full hybrid approach results in sample sizes which are typically much smaller than either previous approach. This is the result of the design prior for the treatment effect $\delta$ giving a high probability to the treatment effect being larger than the MCID. We will investigate the effect of the prior on $\delta$ on the sample size in Section \ref{sec:delta}.

\subsection{The fully Bayesian approach}

We can calculate the required sample size using the assurance based on a Bayesian analysis, as described in Sections \ref{sec:inf} and \ref{sec:ss}. To ensure that an accurate sample size is chosen, we use the analysis priors, numbers of inner loop and outer loop samples and repetitions recommended in Section \ref{sec:comp}. The design priors are identical to those used for the full hybrid method above. The full Bayesian approach defines a design prior distribution over $\bm p$, representing the variability between cluster sizes, rather than over $\nu$. To be consistent with the prior mean and range of $\nu$, we give $\bm p$ a Dirichlet prior distribution with $a_1=\ldots=a_J=7$.

The resulting sample sizes are given in the fourth row of Table \ref{tab:ss}. We see that the smallest sample size, 192 individuals split into 48 clusters, and largest sample size, 230 individuals in 46 clusters, are both smaller than the smallest and largest sample sizes from any of the other methods respectively, including the full hybrid method which used an identical design prior distribution. This implies that basing the success criterion for the trial on the analyis posterior distribution concludes that the treatment is effective more quickly than the equivalent hypothesis test in this case.

\subsection{The effect of the design prior on the treatment effect}
\label{sec:delta} 

The sample sizes from the full Bayesian approach reported in Table \ref{tab:ss} were based on a design prior distribution for the treatment effect of $\delta\sim\textrm{N}(3.5,0.9^2)$. This gave a probability of 0.75 that the treatment effect was above the MCID. Note that this is for the {\em design} of the trial: in the analysis prior this probability is close to zero. In this section we will investigate changes to the mean and standard deviation of this prior, and their effect on the sample size required for the trial.

\begin{figure}[ht]
\centering
\includegraphics[height=7.5cm]{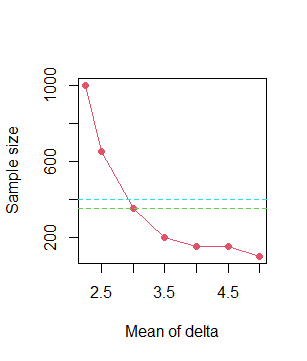}
\includegraphics[height=7.5cm]{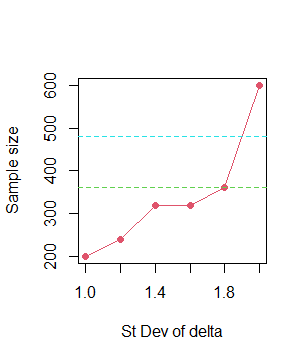}
\includegraphics[height=7.5cm]{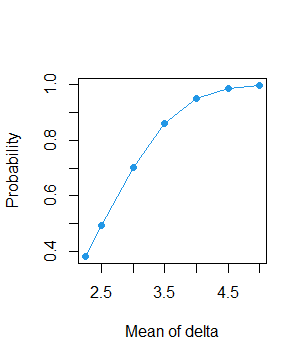}
\includegraphics[height=7.5cm]{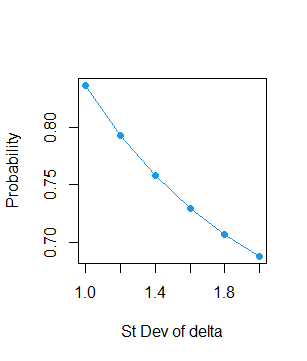}
\caption{Top: The effect of changing the design prior mean (left) and standard deviation (right) for $\delta$ on the sample size required for 50 clusters, based on the fully Bayesian method (red), compared to the sampe sizes from a power calculation (green, dashed) and the hybrid MCID approach (light blue, dashed). Bottom: the probability, under this design prior, that $\delta>2.52$.}
\label{fig:delta}
\end{figure}

We vary the mean of the design prior for $\delta$ in the range $[2.25,5]$ and the standard deviation in the range $[1,2]$. These ranges are chosen to be those in which interesting changes in the sample size required occur. The resulting sample sizes are plotted in the top row of Figure \ref{fig:delta}, together with the probability, under the design prior, that $\delta>2.52$, in the bottom row. We focus on a trial with 50 clusters, the maximum considered in ICONS.

We see that in both cases as the probability, under the design prior, that the treatment effect is larger than the MCID increases, the required sample size decreases. The sample size required decreases as the expected (mean) treatment effect increases, whereas it increases with increasing prior standard deviation on the treatment effect. When the expected treatment effect is at least 3.5 (based on a prior standard deviation of 0.9), the full Bayesian approach requires a lower sample size than a power calculation and the hybrid MCID approach. Similarly, with a prior standard deviation of less than 1.8 (when the prior mean is 3.5), the full Bayesian approach requires a sample size smaller than the power calculation and hybrid MCID approach.

\section{Summary}

In this paper we have considered the choice of sample size, and resulting inference, for a two-arm superiority cluster RCT with continuous outcomes from a Bayesian perspective. We outlined the inference for such a trial based on the posterior distribution for the treatment effect, as previously detailed in \cite{Spi01}. We then developed the form of the assurance, based on this inference, to choose a sample size for the RCT. This required a two loop Monte Carlo approach, sampling from the design prior distribution in the outer loop and then performing MCMC updates based on the analysis prior to obtain samples in the inner loop.

The result was a fully-developed approach which could be used to find a sample size, but for which choices, such as the form of the analysis prior distributions and the number of inner loop and outer loop samples to take to ensure an accurate sample size, were required. Therefore we investigated the effect of different analyis prior distributions and the number of outer loop MC samples and inner loop MCMC samples on the assurance calculation. This led to a recommendation of conjugate inverse gamma analysis prior distributions on the model variances with hyperparameter values of 0.1 and $L=1000$ and $K=1000$ outer loop and inner loop samples for accurate but efficient sample size determination. We also found that repeating the sample size calculation 35 times, and taking the modal sample size, resulted in a high probability of choosing the correct sample size.

Finally, we applied our approach to the design of the ICONS trial. We compared the fully Bayesian approach we have developed to the standard power calculation and two hybrid approaches which used a Bayesian design via assurance and performed inference via the same hypothesis test as the power calculation. We saw that, when we have prior information that suggests the probability that the treatment effect is larger than the MCID is high, then assurance taking this into account, either via the hybrid approach or the fuly Bayesian approach, can reduce the sample size required. In addition, under certain circumstances, using a Bayesian analysis in addition to assurance for the design can lead to further reductions in the required sample size.

\bibliographystyle{plainnat}
\bibliography{refs}

\begin{thebibliography}{14}
\providecommand{\natexlab}[1]{#1}
\providecommand{\url}[1]{\texttt{#1}}
\expandafter\ifx\csname urlstyle\endcsname\relax
  \providecommand{\doi}[1]{doi: #1}\else
  \providecommand{\doi}{doi: \begingroup \urlstyle{rm}\Url}\fi

\bibitem[Clark and Bachmann(2010)]{Cla10}
A.B. Clark and M.O. Bachmann.
\newblock Bayesian methods of analysis for cluster randomized trials with count
  outcome data.
\newblock \emph{Statistics in Medicine}, 29:\penalty0 199--209, 2010.

\bibitem[Eldridge et~al.(2006)Eldridge, Ashby, and Kerry]{Eld06}
S.E. Eldridge, D.~Ashby, and S.~Kerry.
\newblock Sample size for cluster randomized trials: effect of coefficient of
  variation of cluster size and analysis method.
\newblock \emph{International Journal of Epidemiology}, 35:\penalty0
  1292–1300, 2006.

\bibitem[Jones et~al.(2021)Jones, Streeter, and Baker]{Jon21}
B.G. Jones, A.J. Streeter, and A.~Baker.
\newblock Bayesian statistics in the design and analysis of cluster randomised
  controlled trials and their reporting quality: a methodological systematic
  review.
\newblock \emph{Syst Rev}, 10, 2021.

\bibitem[Kikuchi and Gittins(2010)]{Kik10}
T.~Kikuchi and J.~Gittins.
\newblock A behavioural bayes approach for sample size determination in cluster
  randomized clinical trials.
\newblock \emph{Journal of the Royal Statistical Society: Series C (Applied
  Statistics)}, 59\penalty0 (5):\penalty0 875--888, 2010.

\bibitem[Kunzmann et~al.(2021)Kunzmann, Grayling, Lee, Robertson, Rufibach, and
  Wason]{Kun21}
K.~Kunzmann, M.J. Grayling, K.M. Lee, D.S. Robertson, K.~Rufibach, and J.M.S.
  Wason.
\newblock A review of {B}ayesian perspectives on sample size derivation for
  confirmatory trials.
\newblock \emph{The American Statistician}, 75:\penalty0 424--432, 2021.

\bibitem[O'Hagan and Stevens(2001)]{Oha01}
A.~O'Hagan and J.W. Stevens.
\newblock Bayesian assessment of sample size for clinical trials of
  cost-effectiveness.
\newblock \emph{Medical Decision Making}, 21:\penalty0 219--230, 2001.

\bibitem[O'Hagan et~al.(2005)O'Hagan, Stevens, and Campbell]{Oha05}
A.~O'Hagan, J.W. Stevens, and M.J. Campbell.
\newblock Assurance in clinical trial design.
\newblock \emph{Pharmaceutical Statistics}, 4:\penalty0 187--201, 2005.

\bibitem[Plummer(2022)]{Plu22}
M.~Plummer.
\newblock \emph{rjags: Bayesian Graphical Models using MCMC}, 2022.
\newblock URL \url{https://CRAN.R-project.org/package=rjags}.
\newblock R package version 4-13.

\bibitem[Ryan et~al.(2015)Ryan, Drovandi, and Pettitt]{Rya15}
E.G. Ryan, C.C. Drovandi, and A.N. Pettitt.
\newblock Fully bayesian experimental design for pharmacokinetic studies.
\newblock \emph{Entropy}, 17\penalty0 (3):\penalty0 1063--1089, 2015.

\bibitem[Spiegelhalter(2001)]{Spi01}
D.J. Spiegelhalter.
\newblock Bayesian methods for cluster randomized trials with continuous
  responses.
\newblock \emph{Statistics in Medicine}, 20\penalty0 (3):\penalty0 435--452,
  2001.

\bibitem[Thomas et~al.(2015)Thomas, French, Sutton, Forshaw, Leathley, Burton,
  and et~al]{Tho15}
L.H. Thomas, B.~French, C.J. Sutton, D.~Forshaw, M.J. Leathley, C.R. Burton,
  and et~al.
\newblock {I}dentifying {C}ontinence {O}ptio{N}s after {S}troke \text{(ICONS)}:
  an evidence synthesis, case study and exploratory cluster randomised
  controlled trial of the introduction of a systematic voiding programme for
  patients with urinary incontinence after stroke in secondary care.
\newblock \emph{Programme Grants for Applied Research. Southampton (UK): NIHR
  Journals Library}, 2015.

\bibitem[Tishkovskaya et~al.(2022)Tishkovskaya, Sutton, Thomas, and
  Watkins]{Tis22}
S.V. Tishkovskaya, C.J. Sutton, L.H. Thomas, and C.L. Watkins.
\newblock Determining the sample size for a cluster-randomised trial: Bayesian
  hierarchical modelling of the intracluster correlation coefficient.
\newblock \emph{Clinical Trials}, to appear, 2022.

\bibitem[Turner et~al.(2006)Turner, Omar, and S.G.]{Tur06}
R.M. Turner, R.Z. Omar, and Thompson S.G.
\newblock Constructing intervals for the intracluster correlation coefficient
  using {B}ayesian modelling, and application in cluster randomized trials.
\newblock \emph{Statistics in Medicine}, 25:\penalty0 1443--1456, 2006.

\bibitem[Wilson et~al.(2022)Wilson, Williamson, and Tishkovskaya]{Wil22}
K.J. Wilson, S.F. Williamson, and S.V. Tishkovskaya.
\newblock Hybrid sample size calculations for cluster randomised trials using
  assurance.
\newblock \emph{ArXiV}, 2022.

\end{thebibliography}

\end{document}